# Complexity Results in Graph Reconstruction[*]

Edith Hemaspaandra[†]
Department of Computer Science
Rochester Institute of Technology
Rochester, NY 14623, USA

Lane A. Hemaspaandra
Department of Computer Science
University of Rochester
Rochester, NY 14627, USA

Stanisław P. Radziszowski
Department of Computer Science
Rochester Institute of Technology
Rochester, NY 14623, USA

Rahul Tripathi
Department of Computer Science
University of Rochester
Rochester, NY 14627, USA

October 10, 2004

**Abstract**

We investigate the relative complexity of the graph isomorphism problem (GI) and problems related to the reconstruction of a graph from its vertex-deleted or edge-deleted subgraphs (in particular, deck checking (DC) and legitimate deck (LD) problems). We show that these problems are closely related for all amounts $c \geq 1$ of deletion:

1. GI $\equiv_{iso}^{l}$ VDC$_c$, GI $\equiv_{iso}^{l}$ EDC$_c$, GI $\leq_{m}^{l}$ LVD$_c$, and GI $\equiv_{iso}^{p}$ LED$_c$.

2. For all $k \geq 2$, GI $\equiv_{iso}^{p}$ $k$-VDC$_c$ and GI $\equiv_{iso}^{p}$ $k$-EDC$_c$.

3. For all $k \geq 2$, GI $\leq_{m}^{l}$ $k$-LVD$_c$.

4. GI $\equiv_{iso}^{p}$ 2-LVD$_c$.

5. For all $k \geq 2$, GI $\equiv_{iso}^{p}$ $k$-LED$_c$.

For many of these results, even the $c = 1$ case was not previously known.

Similar to the definition of reconstruction numbers $vrn_\exists(G)$ [HP85] and $ern_\exists(G)$ (see p. 120 of [LS03]), we introduce two new graph parameters, $vrn_\forall(G)$ and $ern_\forall(G)$, and give an example of a family $\{G_n\}_{n \geq 4}$ of graphs on $n$ vertices for which $vrn_\exists(G_n) < vrn_\forall(G_n)$. For every $k \geq 2$ and $n \geq 1$, we show that there exists a collection of $k$ graphs on $(2^{k-1} + 1)n + k$ vertices with $2^n$ 1-vertex-preimages, i.e., one has families of graph collections whose number of 1-vertex-preimages is huge relative to the size of the graphs involved.

---

[*]Supported in part by grants NSF-CCR-0311021 and NSF-CCF-0426761. A preliminary version of this paper was presented at the MFCS '04 conference. This report also appears as URCS-TR-2004-852.

[†]Work done in part while on sabbatical at the University of Rochester.

# 1 Introduction

## 1.1 Background

The general form of a combinatorial reconstruction problem is the following: Given a mathematical structure $\mathcal{S}$ and a collection $\mathcal{D}(\mathcal{S})$ of substructures associated with it, is it possible to reconstruct $\mathcal{S}$ (perhaps give or take some natural notion of isomorphism) from $\mathcal{D}(\mathcal{S})$ with some minor imperfections or no imperfections? Such reconstruction problems are interesting not only from a mathematical point of view but also due to their applicability in diverse fields. In bioinformatics, the multiple sequence alignment problem [CL88] is to reconstruct a sequence with minimum gap insertion and maximum number of matching symbols, given a list of protein or DNA sequences. In computer networking, a reconstruction problem can appear in the following scenario: Given a collection of sketches depicting partial network connections in a city from different locations, reconstruct the network of the entire city.

In this paper, we are concerned with reconstruction problems arising in graph theory. The foremost open problems in the theory of reconstruction of graphs are the Reconstruction Conjecture and the Edge-Reconstruction Conjecture. The Reconstruction Conjecture, formulated by Kelly and Ulam in 1942 [Kel42,Ula60], asserts that every finite, simple, undirected graph on at least three vertices is determined uniquely (up to isomorphism— we treat our graphs broadly as unlabeled) by its collection of 1-vertex-deleted subgraphs. Harary [Har64] formulated the Edge-Reconstruction Conjecture, which states that a finite simple graph with at least four edges can be reconstructed from its collection of 1-edge-deleted subgraphs. For more on these conjectures, the reader can refer to a number of excellent survey papers (e.g., [BH77b,Nas78,Man88,Bon91]) and the book by Lauri and Scapellato [LS03].

Nash-Williams [Nas78] posed an interesting, potentially computational problem related to the Reconstruction Conjecture: Given a collection of graphs, how can we decide whether this has been generated from some graph by deleting one vertex every possible way, i.e., whether the collection is *legitimate*? A similar problem has been posed (see, e.g., [Nas78, Man82]) where one asks whether the collection is generated from some graph by deleting one edge every possible way. These problems are known as the *Legitimate Vertex-Deck Problem* (LVD) and the *Legitimate Edge-Deck Problem* (LED). Other, seemingly easier, problems are the *Vertex-Deck Checking Problem* (VDC) and the *Edge-Deck Checking Problem* (EDC). In these, given a graph $G$ and a collection $\mathcal{D}$ of graphs, we ask whether $\mathcal{D}$ can be generated from $G$ by deleting one vertex, respectively one edge, every possible way.

Mansfield [Man82] and Kratsch and Hemaspaandra [KH94] studied complexity aspects of legitimate deck problems and deck checking problems. Mansfield [Man82] showed that LVD is polynomial-time many-one hard for the Graph Isomorphism problem (which we will often refer to as GI) and that LED is polynomial-time Turing equivalent to the Graph Isomorphism problem. Kratsch and Hemaspaandra [KH94] showed that LVD is logspace many-one hard for the Graph Isomorphism problem, proved that GI is logspace isomorphic to VDC, and obtained polynomial-time algorithms for LVD when restricted to certain classes of graphs—including graphs of bounded degree, partial $k$-trees for any fixed $k$, and graphs of bounded genus. Köbler, Schöning, and Torán [KST93] showed that if the Reconstruction Conjecture holds then LVD is in the complexity class LWPP. And so, conditional on the truth of the Reconstruction Conjecture, Köbler, Schöning, and Torán showed that LVD is low for PP, i.e., $PP^{LVD} = PP$. This result can be viewed as suggesting that LVD cannot be NP-complete, since if it were NP-complete, then the abovementioned LWPP result would immediately imply that either the Reconstruction Conjecture fails or $PP^{NP} = PP$. However, both these claims are widely suspected to be false.

## 1.2 Our Contributions

A more general reconstruction problem deals with collections consisting of all subgraphs obtained through the deletion of (exactly) some fixed number $c \geq 1$ of vertices (or edges). Kelly [Kel57] was the first to look in this direction, Manvel [Man74] made some observations on this problem, and Bondy [Bon91, Section 11.2] surveyed related results. See also Nýdl's review [Nýd01] of the progress made on this problem in the past three decades. In this paper, one of our investigations is of the complexity of legitimate deck problems and deck checking problems for the general case of deletion of some fixed number $c \geq 1$ of vertices (or edges) of a graph. We observe that the logspace isomorphism known to hold between GI and VDC [KH94] also holds, for every $c \geq 1$, between GI and $VDC_c$ and between GI and $EDC_c$. Here and henceforward, the subscript "$c$" in the name of a problem refers to the more general problem based on the deletion of $c$ vertices or edges of a graph. We strengthen the result of Mansfield [Man82] to show that, for every $c \geq 1$, GI is polynomial-time isomorphic to $LED_c$. For $LVD_c$, we observe that for every $c \geq 1$, GI $\leq_m^p LVD_c$ (the $c = 1$ case of this already follows from a result of Kratsch and Hemaspaandra [KH94]). These results appear in Section 3.1.

We next look at the question of reconstructing a graph from a subdeck (a subset of all possible vertex-deleted or edge-deleted subgraphs). See [HP66,Bon69,Lau83] for discussion of this line of investigation in the reconstruction of trees. Our results on the complexity aspects of the reconstruction of a graph from a subdeck are described in Section 3.2. We again show a strong relationship between these problems and the graph isomorphism problem.

Harary and Plantholt [HP85] introduced a parameter, called the ally-reconstruction number of a graph $G$ (which we will denote $vrn_\exists(G)$), and defined it as the minimum number of 1-vertex-deleted subgraphs needed to identify $G$ (as always, up to isomorphism). A similar definition is used for the reconstruction number $ern_\exists(G)$, which is defined in terms of 1-edge-deleted subgraphs (see p. 120 of [LS03]). We introduce two new parameters, $vrn_\forall(G)$ and $ern_\forall(G)$, for a graph $G$, and we give an example of a family $\{G_n\}_{n\geq 4}$ of graphs on $n$ vertices for which $vrn_\exists(G_n) < vrn_\forall(G_n)$. We also give a family of collections of $k$ graphs on $(2^{k-1}+1)n+k$ vertices with $2^n$ 1-vertex-preimages, thus constructing an exponential richness of preimages. These results appear in Section 4.

## 2 Preliminaries

### 2.1 Notation

Our alphabet is $\Sigma = \{0,1\}$. We use $\{.,\ldots,.\}$ to denote sets and $[.,\ldots,.]$ to denote multisets. We use $\cup$ to denote set union as well as multiset union. Let $\langle\ldots\rangle$ be a multi-arity, polynomial-time computable, and polynomial-time invertible pairing function (e.g., that of [HHT97]). We tacitly assume that multisets and graphs are encoded in a standard fashion. For background in complexity theory and for notions such as P, NP, reductions, and completeness, we refer the reader to any book on complexity theory, for example [HO02]. We consider only finite, undirected graphs with no self-loops. Given a graph $G$, let $V(G)$ denote its vertex set and let $E(G)$ denote its edge set. For notational convenience, we sometimes represent a graph $G$ by $(V,E)$, where $V = V(G)$ and $E = E(G)$. By the order of a graph $G$ we mean $||V(G)||$, i.e., the cardinality of its vertex set. The degree of a vertex $v$ in $G$, denoted by $deg_G(v)$, is the number of edges incident on $v$. $\delta(G) = \min\{deg_G(v) \mid v \in V(G)\}$ and $\lambda(G)$ is the minimum number of edges whose deletion from $G$ disconnects $G$. The closed neighborhood $N_G(v)$ of a vertex $v$ in a graph $G$ is the set of vertices that are at distance at most one from $v$, that is, $N_G(v) = \{v\} \cup \{w \mid \{v,w\} \in E(G)\}$. The notions of union and join of graphs here will always implicitly require disjoint sets of vertices and thus for graphs $G$ and $H$ with $V(G) \cap V(H) \neq \emptyset$, we assume that isomorphs $\widehat{G}$ and $\widehat{H}$ of $G$ and $H$, with $V(\widehat{G}) \cap V(\widehat{H}) = \emptyset$, are used in place of $G$ and $H$. The union of graphs $G_1, G_2, \ldots, G_k$, $k \geq 2$, is denoted by $G = G_1 \cup G_2 \cup \cdots \cup G_k$, where $V(G) = \bigcup_{i=1}^k V(G_i)$ and $E(G) = \bigcup_{i=1}^k E(G_i)$.

For a graph $G$ and an integer $m \geq 1$, $mG$ represents the union of $m$ vertex-disjoint (isomorphic) copies of $G$. The join of graphs $G_1, \ldots, G_k$, $k \geq 2$, is denoted by $G = G_1 + \ldots + G_k$, where $V(G) = \bigcup_{i=1}^{k} V(G_i)$ and $E(G) = \bigcup_{i=1}^{k} E(G_i) \cup \bigcup_{i \neq j} \{\{u, v\} \mid u \in V(G_i) \wedge v \in V(G_j)\}$.

For $n \geq 1$, $K_n$ is the complete graph on $n$ vertices and $P_n$ is the path graph on $n$ vertices, i.e., $(V(P_n), E(P_n)) = (\{1, \ldots, n\}, \{\{i, i+1\} \mid 1 \leq i \leq n-1\})$. The line graph $L(G)$ of a graph $G$ is defined by: $V(L(G)) = E(G)$ and $E(L(G)) = \{\{e_1, e_2\} \mid e_1, e_2 \in E(G) \wedge e_1$ and $e_2$ have exactly one vertex in common$\}$. The complement $\overline{G}$ of a graph $G$ is defined by: $V(\overline{G}) = V(G)$ and $E(\overline{G}) = \{\{v, w\} \mid v, w \in V(G), v \neq w, \text{and } \{v, w\} \notin E(G)\}$.

Given a graph $G$ and a set $S \subseteq V(G)$, $G - S$ denotes a graph with $V(G - S) = V(G) - S$ and $E(G - S) = E(G) - \{\{u, v\} \mid \{u, v\} \in E(G) \wedge \{u, v\} \cap S \neq \emptyset\}$. Similarly, if $S \subseteq E(G)$, then $G - S$ denotes a graph with $V(G - S) = V(G)$ and $E(G - S) = E(G) - S$. We will call any collection of graphs with the same number of vertices a "vertex-deck" and will use the term "edge-deck" to denote a collection of graphs with the same number of edges. The graphs in a vertex-deck are called vertex-cards and the graphs in an edge-deck are called edge-cards. For a graph $G$ and for any $c \geq 1$, the $c$-vertex-deleted-deck of $G$, denoted by vertex-deck$_c(G)$, is the multiset $[G - S \mid S \subseteq V(G)$ and $||S|| = c]$, and the $c$-edge-deleted-deck of $G$, denoted by edge-deck$_c(G)$, is the multiset $[G - S \mid S \subseteq E(G)$ and $||S|| = c]$. We say that a vertex-deck $D_1 = [G_1, \ldots, G_n]$ is equivalent to a vertex-deck $D_2 = [G'_1, \ldots, G'_{n'}]$, denoted by $D_1 = D_2$, if $n = n'$ and there exists a one-to-one mapping that maps each graph from $D_1$ to an isomorphic graph from $D_2$. We use an analogous definition for the equivalence of two edge-decks: An edge-deck $D_1 = [G_1, \ldots, G_m]$ is equivalent to an edge-deck $D_2 = [G'_1, \ldots, G'_{m'}]$, denoted by $D_1 = D_2$, if $m = m'$ and there exists a one-to-one mapping that maps each graph from $D_1$ to an isomorphic graph from $D_2$. The notion of $D_1 \subseteq D_2$ is defined analogously. For any $c \geq 1$, we say a graph $G$ is a $c$-vertex-preimage of $[G_1, \ldots, G_k]$ if $[G_1, \ldots, G_k] \subseteq$ vertex-deck$_c(G)$, and we say a graph $G$ is a $c$-edge-preimage of $[G_1, \ldots, G_k]$ if $[G_1, \ldots, G_k] \subseteq$ edge-deck$_c(G)$. The reason these definitions have "$\subseteq$"s rather than "="s is so that our notions of preimage apply meaningfully both to (full) decks and to "subdecks." Typically, when we are speaking of preimages of (full) decks, the number of vertices (or edges, in the case of edge-preimages) will make it clear that this is the case. However, we will at times use the terms $c$-vertex-pure-preimage and $c$-edge-pure-preimage when we wish to specifically emphasize the equality case: For any $c \geq 1$, we say a graph $G$ is a $c$-vertex-pure-preimage of $[G_1, \ldots, G_k]$ if $[G_1, \ldots, G_k] =$ vertex-deck$_c(G)$, and similarly for the edge case. For any $c \geq 1$, we say that a graph $H$ is a $c$-vertex-card ($c$-edge-card) of a graph $G$ if $H$ is isomorphic to a graph in vertex-deck$_c(G)$ (edge-deck$_c(G)$).

$[G_1, \ldots, G_k]$ is a legitimate $c$-vertex-deck ($c$-vertex-subdeck) if there is a graph $G$ such that $[G_1, \ldots, G_k] = \text{vertex-deck}_c(G)$ ($[G_1, \ldots, G_k] \subseteq \text{vertex-deck}_c(G)$). The notions of legitimate $c$-edge-deck and legitimate $c$-edge-subdeck, for any $c \geq 1$, are defined analogously. For any graph $G$, the endvertex-deck of $G$, denoted by endvertex-deck$(G)$, is the multiset consisting of the subgraphs $G - v$ where $v$ is an endvertex of $G$, i.e., a vertex for which $deg_G(v) = 1$.

## 2.2 Graph Isomorphism

A graph $G$ is isomorphic to a graph $H$ if there is a bijective mapping $\psi : V(G) \to V(H)$ such that, for all $v_1, v_2 \in V(G)$, $\{v_1, v_2\} \in E(G)$ if and only if $\{\psi(v_1), \psi(v_2)\} \in E(H)$. In this case, $\psi$ is called an isomorphism between graphs $G$ and $H$, and we write $G \cong H$. The graph isomorphism problem, GI, is $\{\langle G, H \rangle \mid G \cong H\}$. Here and in other such cases, we are viewing encoding and decoding as transparent and implicit. This is not a totally innocuous assumption, since isomorphisms of a problem may be ruined under particularly kinky encodings. However, the natural encodings of the problems used here, for those problems for which we assert isomorphisms, have the type of padding/etc. functions needed to prove isomorphisms (see Section 2.3), so as is typical in papers on isomorphism we do not focus on encoding details.

The graph isomorphism problem is of great interest to mathematicians and theoretical computer scientists. Arvind and Kurur [AK02b] showed recently that GI is in the PP-low complexity class SPP. GI is also known to be in the complexity class NP $\cap$ coAM, which is low for $\Sigma_2^p$ (i.e., $(\Sigma_2^p)^{\text{NP} \cap \text{coAM}} = \Sigma_2^p$, see [GMW91,GS89,Sch87]) and, as established by Arvind and Köbler, even for the class (which we do not define here) ZPP$^{\text{NP}}$ [AK02a]; so the polynomial hierarchy would collapse if GI were NP-complete (or were anywhere in the "high hierarchy," see [Sch83]). These facts support the widely accepted belief that GI is not NP-complete.

**Definition 2.1 ([KST93], see also [Kad88,RR92,LT92])** *An* or-function *for a set $A$ is a function $f$ mapping sequences of strings to strings such that for every sequence $x_1, \ldots, x_n$, $n \geq 1$, it holds that $f(\langle x_1, \ldots, x_n \rangle) \in A \iff (\exists i \in \{1, \ldots, n\})[x_i \in A]$. The* and-function *for a set $A$ is defined analogously.*

**Proposition 2.2 ([KST93])** GI *has a polynomial-time computable or-function and a polynomial-time computable and-function (both of them in the sense of Definition 2.1).*

The fact that GI has a polynomial-time computable or-function immediately implies the following corollary, which will be useful as we seek to obtain polynomial-time many-one reductions from certain sets to GI.

**Corollary 2.3 ([KST93])** *For every set L, if L disjunctive polynomial-time truth-table reduces to GI, then L polynomial-time many-one reduces to GI.*

## 2.3 A Tool for Proving Isomorphism between Sets

**Definition 2.4** *A set A is* logspace (polynomial-time) isomorphic *to a set B, denoted by $A \equiv_{iso}^{l} B$ ($A \equiv_{iso}^{p} B$), if there exists a bijection $f : \Sigma^* \to \Sigma^*$ such that f is a logspace (polynomial-time) many-one reduction from A to B and $f^{-1}$ is a logspace (polynomial-time) reduction from B to A.*

The following results of Berman and Hartmanis [Har78,BH77a] give a sufficient condition for showing logspace or polynomial-time isomorphism between sets. (The wording of them used here mostly follows the presentation of [KH94].) We will use Theorem 2.8 to help us show isomorphism between GI and certain problems considered in this paper.

**Lemma 2.5 ([Har78,BH77a])** *Let A be a set for which logspace (polynomial-time) computable functions $S_A$ and $D_A$ exist such that*

1. *$(\forall x, y)[S_A(x, y) \in A \iff x \in A]$, and*

2. *$(\forall x, y)[D_A(S_A(x, y)) = y]$.*

*If f is any logspace (polynomial-time) reduction from C to A, the mapping $g(x) = S_A(f(x), x)$ is a one-to-one logspace (polynomial-time) reduction from C to A and $g^{-1}$ is logspace (polynomial-time) computable.*

**Definition 2.6 ([Har78,BH77a])** *$Z_A : \Sigma^* \to \Sigma^*$ is a* padding function *for the set A if (a) $Z_A$ is one-to-one, and (b) $(\forall x)[Z_A(x) \in A \iff x \in A]$.*

**Lemma 2.7 ([Har78,BH77a])** *Let f be a one-to-one logspace (polynomial-time) computable reduction from A to B and let $f^{-1}$ be logspace (polynomial-time) computable. Assume that there is a function Z that is a padding function for at least one of A or B, and that has the following properties:*

1. *Z and $Z^{-1}$ are logspace (polynomial-time) computable.*

2. $(\forall x)\left[|Z(x)| > |x|^2 + 1\right]$. $((\forall x)\left[|Z(x)| > |x|\right].)$

Then there exists a one-to-one logspace (polynomial-time) reduction $g$ from $A$ to $B$ such that

1. $g^{-1}$ is logspace (polynomial-time) computable, and

2. $(\forall x)\left[|g(x)| > |x|^2\right]$  $((\forall x)\left[|g(x)| > |x|\right])$.

**Theorem 2.8 ([Har78,BH77a])** *Let $A$ and $B$ be sets such that $A$ is logspace (polynomial-time) reducible to $B$ and $B$ is logspace (polynomial-time) reducible to $A$. Furthermore, let the set $A$ have a logspace (polynomial-time) padding function $Z_A$ satisfying Lemma 2.7 and let $A$ also have logspace (polynomial-time) functions $S_A$ and $D_A$ satisfying Lemma 2.5. Then if $B$ has logspace (polynomial-time) functions $S_B$ and $D_B$ satisfying Lemma 2.5, then $B$ is logspace (polynomial-time) isomorphic to $A$.*

The existence of logspace (and therefore, polynomial-time) computable functions $S_{\text{GI}}$ and $D_{\text{GI}}$ satisfying Lemma 2.5, and $Z_{\text{GI}}$ satisfying Lemma 2.7 is already known. We refer the reader to [Boo78,KH94] for the proofs.

**Lemma 2.9 ([Boo78,KH94])** *GI has logspace functions $S_{\text{GI}}$ and $D_{\text{GI}}$ satisfying Lemma 2.5.*

**Lemma 2.10 ([Boo78,KH94])** *GI has a logspace padding function $Z_{\text{GI}}$ satisfying Lemma 2.7.*

## 2.4 Computational Problems in Graph Reconstruction

Kelly [Kel57] first proposed the idea of generalizing the Reconstruction Conjecture to $c$-vertex-deleted subgraphs for $c > 1$. Kelly showed that there are graphs that are not determined uniquely (up to isomorphism) by their 2-vertex-deleted subgraphs. However, it is suspected that, for any $c > 1$, all sufficiently large graphs satisfy the general reconstruction problem for $c$-vertex-deleted subgraphs. From a computational complexity point of view, it is natural to seek to understand the complexity of problems related to the reconstruction of a graph from its $c$-vertex-deleted or $c$-edge-deleted subgraphs for different values of $c$. With this motivation, we state the computational problems we study in this paper.

1. VERTEX-DECK CHECKING$_c$ (abbreviated VDC$_c$)
   VDC$_c = \{\langle G, [G_1, \ldots, G_n]\rangle \mid [G_1, \ldots, G_n] = \text{vertex-deck}_c(G)\}$.

2. EDGE-DECK CHECKING$_c$ (abbreviated EDC$_c$)
   EDC$_c = \{\langle G, [G_1, \ldots, G_m]\rangle \mid [G_1, \ldots, G_m] = \text{edge-deck}_c(G)\}$.

3. LEGITIMATE VERTEX-DECK$_c$ (abbreviated LVD$_c$)
   LVD$_c = \{[G_1, \ldots, G_n] \mid (\exists G)[[G_1, \ldots, G_n] = \text{vertex-deck}_c(G)]\}$.

4. LEGITIMATE EDGE-DECK$_c$ (abbreviated LED$_c$)
   LED$_c = \{[G_1, \ldots, G_m] \mid (\exists G)[[G_1, \ldots, G_m] = \text{edge-deck}_c(G)]\}$.

For any fixed $k \geq 2$, one can study the $k$ $c$-vertex-(edge-)card versions of the abovementioned problems. In these versions one is given just $k$ cards, allegedly from a deck based on the deletion of $c$ vertices (edges). These problems will be denoted $k$-VDC$_c$, $k$-EDC$_c$, $k$-LVD$_c$, and $k$-LED$_c$, and are defined as follows.

1. $k$-VERTEX-SUBDECK CHECKING$_c$ (abbreviated $k$-VDC$_c$)
   $k$-VDC$_c = \{\langle G, [G_1, \ldots, G_k]\rangle \mid [G_1, \ldots, G_k] \subseteq \text{vertex-deck}_c(G)\}$.

2. $k$-EDGE-SUBDECK CHECKING$_c$ (abbreviated $k$-EDC$_c$)
   $k$-EDC$_c = \{\langle G, [G_1, \ldots, G_k]\rangle \mid [G_1, \ldots, G_k] \subseteq \text{edge-deck}_c(G)\}$.

3. $k$-LEGITIMATE VERTEX-SUBDECK$_c$ (abbreviated $k$-LVD$_c$)
   $k$-LVD$_c = \{[G_1, \ldots, G_k] \mid (\exists G)[[G_1, \ldots, G_k] \subseteq \text{vertex-deck}_c(G)]\}$.

4. $k$-LEGITIMATE EDGE-SUBDECK$_c$ (abbreviated $k$-LED$_c$)
   $k$-LED$_c = \{[G_1, \ldots, G_k] \mid (\exists G)[[G_1, \ldots, G_k] \subseteq \text{edge-deck}_c(G)]\}$.

## 3 Reconstruction from Vertex and Edge Decks

### 3.1 Reconstruction from a Complete Deck

In this section, we investigate the complexity of VDC$_c$, EDC$_c$, LVD$_c$, and LED$_c$, for each $c \geq 1$. Kratsch and Hemaspaandra [KH94] showed that GI is logspace isomorphic to VDC$_1$. We extend this result, and state that, for all $c \geq 1$, GI is logspace isomorphic to VDC$_c$ as well as to EDC$_c$.

**Theorem 3.1**   *1. For all $c \geq 1$, GI is logspace isomorphic to VDC$_c$.*

*2. For all $c \geq 1$, GI is logspace isomorphic to EDC$_c$.*

**Proof** The proofs of both parts follow from the techniques used in [KH94, Lemmas 3.1, 3.2, and 5.6] and so are omitted. ∎

Kratsch and Hemaspaandra [KH94] showed that GI $\leq_m^l$ LVD$_1$. We extend this result and as Theorem 3.2 show that, for any $c \geq 1$, GI $\leq_m^l$ LVD$_c$. Mansfield [Man82] showed that GI is polynomial-time Turing equivalent to LED$_1$. In Theorem 3.5, building on Lemmas 3.3 and 3.4, we extend this result and show that, for each $c \geq 1$, GI is polynomial-time isomorphic to LED$_c$.

**Theorem 3.2** *For all $c \geq 1$, GI $\leq_m^l$ LVD$_c$.*

**Proof** The construction in this proof is inspired by the construction in [KH94, Theorem 4.1]. Without loss of generality, we restrict ourselves to input instances $\langle G, H \rangle$ of GI such that $G$ and $H$ are connected graphs with at least three vertices. The logspace many-one reduction $\sigma$ is defined as follows:

$$\sigma(\langle G, H \rangle) = (\text{vertex-deck}_c(G \cup (c+1)K_1) - [G \cup K_1]) \cup [H \cup K_1].$$

(Just to be clear, note that the three inner unions are over graphs, but the "−" and the outer union both are operations on multisets. We use $[A]$ to coerce a graph $A$ into a singleton multiset.) Clearly, $G \cong H$ implies that $G \cup (c+1)K_1$ is a $c$-vertex-pure-preimage of the vertex-deck $\sigma(\langle G, H \rangle)$.

We now turn to the proof that $(\text{vertex-deck}_c(G \cup (c+1)K_1) - [G \cup K_1]) \cup [H \cup K_1] \in \text{LVD}_c$ implies $G \cong H$. Let $\mathcal{G}$ be a $c$-vertex-pure-preimage of the vertex-deck $\sigma(\langle G, H \rangle)$. Since $c \geq 1$, $G \cup K_1$ is a vertex-card in the $c$-vertex-deck of $\mathcal{G}$. Thus, $\mathcal{G}$ can be obtained from $G \cup K_1$ by adding $c$ vertices and 0 or more edges each incident on at least one of the $c$ added vertices. Suppose that there is an edge $e$ incident on one of the new vertices. If $e$ connects the new vertex to $G$, then there exists a connected vertex-card in the vertex-deck. This is a contradiction, since the above deck clearly has no connected card. So assume that $e$ connects the new vertex either to another of the new vertices or to the isolated vertex in $G \cup K_1$. Then there is a vertex-card in the $c$-vertex-deck that consists of, as its connected components, a connected graph of order $||V(G)|| - 1$ and $K_2$. But clearly there is no such vertex-card in the vertex-deck. It follows that $\mathcal{G} \cong G \cup (c+1)K_1$. The only way that $H \cup K_1$ can be in vertex-deck$_c(\mathcal{G})$ is if $G \cong H$. ∎

Note that Theorem 3.2's proof does not count edges, as is done in the proof of [KH94, Theorem 4.1]. Lemma 3.3 can be proved using the proof method of Theorem 3.2. However, the proof of Theorem 3.5 requires, for any $c \geq 1$, a logspace-invertible reduction function from GI to $\text{LED}_c$. Since the technique used in the proof of Theorem 3.2 does not give any obvious logspace-invertible reduction function, we give a different proof of Lemma 3.3, which does yield one.

**Lemma 3.3** *For all $c \geq 1$, $\text{GI} \leq_m^l \text{LED}_c$.*

**Proof** Fix a $c \geq 1$. Without loss of generality, we restrict ourselves to input instances $\langle G, H \rangle$ of GI such that $G$ and $H$ are connected graphs on $n > \max\{c, 2\}$ vertices. Let $\ell = n + 1$. Define a logspace-computable function $\sigma$ as follows:

$$\sigma(\langle G, H \rangle) = $$
$$[H \cup 2cK_1 \cup K_\ell] \cup (\text{edge-deck}_c(G \cup cK_2 \cup K_\ell) - [G \cup 2cK_1 \cup K_\ell]).$$

Clearly, if $G \cong H$, then $G \cup cK_2 \cup K_\ell$ is a $c$-edge-pure-preimage of the edge-deck $\sigma(\langle G, H \rangle)$. We now prove the converse. Suppose that $\sigma(\langle G, H \rangle) \in \text{LED}_c$. Let $e_1, \ldots, e_c$ be $c$ edges in $K_\ell$. A $c$-edge-preimage $\mathcal{G}$ of the edge-deck $\sigma(\langle G, H \rangle)$ can be obtained by adding $c$ edges to the edge-card $G \cup cK_2 \cup (K_\ell - \{e_1, \ldots, e_c\})$. Note that $K_\ell$ is a subgraph of $\mathcal{G}$, since $H \cup 2cK_1 \cup K_\ell$ is an edge-card. The only way for $\mathcal{G}$ to include $K_\ell$ as a subgraph is to add the $c$ edges $e_1, \ldots, e_c$ to $K_\ell - \{e_1, \ldots, e_c\}$, since it will require more than $c$ edges to form a complete graph $K_\ell$ that includes any vertex from $G \cup cK_2$. Thus, $G \cup cK_2 \cup K_\ell$ is the unique $c$-edge-preimage (up to isomorphism) of $\sigma(\langle G, H \rangle)$. $H \cup 2cK_1 \cup K_\ell$ is a card in the $c$-edge-deck of $\mathcal{G}$ and the only way to turn $H \cup 2cK_1 \cup K_\ell$ into the $c$-edge-preimage is to add $c$ edges to $2cK_1$, since the $c$-edge-preimage has no isolated vertices. Thus, $G$ must be isomorphic to $H$. ∎

**Lemma 3.4** *For all $c \geq 1$, $\text{LED}_c \leq_m^p \text{GI}$.*

**Proof** Fix a $c \geq 1$. We first show that $\text{LED}_c \leq_{dtt}^p \text{EDC}_c$. Let $[G_1, \ldots, G_m]$ be an instance of $\text{LED}_c$. By definition,

$$[G_1, \ldots, G_m] \in \text{LED}_c \iff$$
$$\bigvee_{\widehat{E} \subseteq E(\overline{G_1}),\ ||\widehat{E}||=c} \langle (V(G_1), E(G_1) \cup \widehat{E}), [G_1, \ldots, G_m] \rangle \in \text{EDC}_c.$$

This shows that $\text{LED}_c \leq_{dtt}^p \text{EDC}_c$. Since $\text{EDC}_c \leq_m^p \text{GI}$ (Theorem 3.1(2)), $\text{LED}_c \leq_{dtt}^p \text{GI}$. By Corollary 2.3, it follows that $\text{LED}_c \leq_m^p \text{GI}$. ∎

**Theorem 3.5** *For every $c \geq 1$, GI is polynomial-time isomorphic to $\text{LED}_c$.*

**Proof** Fix a $c \geq 1$. GI $\equiv_m^p$ LED$_c$ follows from Lemmas 3.3 and 3.4. By Theorem 2.8 and Lemmas 2.9 and 2.10, it suffices to show that LED$_c$ has polynomial-time computable functions $S_{\text{LED}_c}$ and $D_{\text{LED}_c}$ satisfying Lemma 2.5. The function $S_{\text{LED}_c}$ is defined as follows: On input $([G_1, \ldots, G_m], y)$,

1. Compute the polynomial-time many-one reduction from LED$_c$ to GI on input $[G_1, \ldots, G_m]$. Let $\langle \mathcal{H}_1, \mathcal{H}_2 \rangle$ be the output of the reduction.

2. Compute $S_{\text{GI}}(\langle \mathcal{H}_1, \mathcal{H}_2 \rangle, y)$ and let $\langle \widehat{\mathcal{H}}_1, \widehat{\mathcal{H}}_2 \rangle$ be the output of this step.

3. Compute the logspace many-one reduction $\sigma$ from GI to LED$_c$ (defined in Lemma 3.3) on input $\langle \widehat{\mathcal{H}}_1 + K_{c+2}, \widehat{\mathcal{H}}_2 + K_{c+2} \rangle$ and output the string computed in this step.

From the definition of many-one reducibility and that of $S_{\text{GI}}$, it follows that $S_{\text{LED}_c}([G_1, \ldots, G_m], y) \in \text{LED}_c \iff [G_1, \ldots, G_m] \in \text{LED}_c$. We now define the function $D_{\text{LED}_c}$ as follows. On input $[H_1, \ldots, H_m]$,

1. Scan $[H_1, \ldots, H_m]$ to find an edge-card of the form $(\widehat{\mathcal{H}}_1 + K_{c+2}) \cup cK_2 \cup (K_\ell - \{e_i \mid 1 \leq i \leq c \text{ and } e_i \in E(K_\ell)\})$, where $\ell = \|V(\widehat{\mathcal{H}}_1)\| + c + 3$. If no such card exists, then output the string "undefined." (To see that this step can be done in polynomial time, it may be helpful to keep in mind that, for each fixed $d$, the number of $d$-cliques a graph can have is polynomial in the size of the graph, and one can quickly enumerate all such.)

2. Scan $[H_1, \ldots, H_m]$ to find an edge-card of the form $(\widehat{\mathcal{H}}_2 + K_{c+2}) \cup 2cK_1 \cup K_\ell$, where $\ell = \|V(\widehat{\mathcal{H}}_2)\| + c + 3$. If no such card exists, then output the string "undefined."

3. Output $D_{\text{GI}}(\langle \widehat{\mathcal{H}}_1, \widehat{\mathcal{H}}_2 \rangle)$.

By the construction of the many-one reduction in Lemma 3.3 and the definition of $D_{\text{GI}}$, it is easy to see that $D_{\text{LED}_c}$ satisfies Lemma 2.5. ∎

## 3.2 Reconstruction from a Subdeck

In this section, we investigate the complexity of problems related to the reconstruction of a graph from its partial (incomplete) deck of vertex-deleted or edge-delete subgraphs. We first show, in Lemma 3.6, that there is a close connection between the $c$-edge-deleted-deck of $G$ and the $c$-vertex-deleted-deck of $L(G)$ (the line graph of $G$).

**Lemma 3.6** *For all $c \geq 1$, and for all graphs $G$, if* edge-deck$_c(G) = [G_1, \ldots, G_m]$, *then*

$$\text{vertex-deck}_c(L(G)) = [L(G_1), \ldots, L(G_m)].$$

**Proof** Let $G = (V, E)$. By definition, $L(G) = (E, \widehat{E})$, where $\widehat{E} = \{\{e_1, e_2\} \mid e_1, e_2 \in E$, and $e_1$ and $e_2$ share exactly one vertex$\}$, edge-deck$_c(G) = [G - E' \mid E' \subseteq E$ and $||E'|| = c]$, and vertex-deck$_c(L(G)) = [L(G) - E' \mid E' \subseteq E$ and $||E'|| = c]$.

To prove the lemma, it suffices to show that for all $E' \subseteq E$ such that $||E'|| = c$, $L(G - E') \cong L(G) - E'$. This is easy to see, since $L((V, E - E')) = (E - E', \widehat{E'})$, where $\widehat{E'} = \{\{e_1, e_2\} \mid e_1, e_2 \in E - E'$, and $e_1$ and $e_2$ share exactly one vertex$\}$. ∎

Hemminger [Hem69] proved a much stronger result than the one stated in Lemma 3.6 for the case of $c = 1$.

**Theorem 3.7 ([Hem69])** *For any graph $G$, $G$ can be determined uniquely up to isomorphism from* edge-deck$_1(G)$ *if and only if $L(G)$ can be determined uniquely up to isomorphism from* vertex-deck$_1(L(G))$.

However, for our proofs, we only need the result stated in Lemma 3.6.

In Lemma 3.9, we show that, for any $k \geq 2$, under certain restrictions on graph $G$, a relationship similar to Lemma 3.6 holds between a collection of $k$ edge-cards and the corresponding collection of $k$ vertex-cards. We use Lemma 3.9 in proving Lemma 3.10, which states that for any $c \geq 1$ and $k \geq 2$, $k$-EDC$_c \leq^l_m k$-VDC$_c$. In proving Lemma 3.9, we use the following theorem by Whitney [Whi32].

**Theorem 3.8 ([Whi32], see also [Har71])** *If $G$ and $H$ are connected graphs other than $K_3$, then $G \cong H$ if and only if $L(G) \cong L(H)$.*

**Lemma 3.9** *For all $c \geq 1$ and $k \geq 2$, for all graphs $G$ with $n \geq 4$ vertices and edge-connectivity $\lambda(G) > c$, and for all connected graphs $G_1, \ldots, G_k$ with $n$ vertices,*

$$[G_1, \ldots, G_k] \subseteq \text{edge-deck}_c(G) \iff [L(G_1), \ldots, L(G_k)] \subseteq \text{vertex-deck}_c(L(G)).$$

**Proof** Note that the left-to-right direction follows immediately from Lemma 3.6. For the converse, suppose that $G_1, \ldots, G_k$ are connected graphs with $n$ vertices such that $[L(G_1), \ldots, L(G_k)] \subseteq \text{vertex-deck}_c(L(G))$. By Lemma 3.6, there exist $H_1, \ldots, H_k$ such that $[H_1, \ldots, H_k] \subseteq \text{edge-deck}_c(G)$ and $[L(H_1), \ldots, L(H_k)] = [L(G_1), \ldots, L(G_k)]$. Since $\lambda(G) > c$ and $n \geq 4$, each of the graphs $H_1, \ldots, H_k$ is a connected graph other than $K_3$. It is also true that each $G_i$ is a connected graph other than $K_3$ (since $n \geq 4$). It follows by Theorem 3.8 that $[H_1, \ldots, H_k] = [G_1, \ldots, G_k]$. Thus, $[G_1, \ldots, G_k] \subseteq \text{edge-deck}_c(G)$. ∎

### 3.2.1 Subdeck Checking Problems

**Lemma 3.10** *For all $c \geq 1$ and $k \geq 2$, $k\text{-EDC}_c \leq_m^l k\text{-VDC}_c$.*

**Proof** Fix a $c \geq 1$ and a $k \geq 2$. For $H$ a graph on $n$ vertices, define $\widehat{H}$ as $H + (K_{n+1} \cup \{v_0\})$.

Let $\langle G, [G_1, \ldots, G_k] \rangle$ be an instance of $k\text{-EDC}_c$. W.l.o.g, we assume that $G, G_1, \ldots, G_k$ are graphs on $n > c$ vertices. The logspace many-one reduction $\sigma$ from $k\text{-EDC}_c$ to $k\text{-VDC}_c$ is defined by

$$\sigma(\langle G, [G_1, \ldots, G_k] \rangle) = \langle L(\widehat{G}), [L(\widehat{G}_1), \ldots, L(\widehat{G}_k)] \rangle.$$

Clearly, $\sigma$ is computable in logspace. To prove that $\sigma$ is a many-one reduction from $k\text{-EDC}_c$ to $k\text{-VDC}_c$, we first show that $\langle G, [G_1, \ldots, G_k] \rangle \in k\text{-EDC}_c$ if and only if $\langle \widehat{G}, [\widehat{G}_1, \ldots, \widehat{G}_k] \rangle \in k\text{-EDC}_c$. The left-to-right direction is immediate. For the converse, suppose that there exist $k$ distinct sets of $c$ edges $E_1, \ldots, E_k$ in $\widehat{G}$ such that $\widehat{G} - E_i \cong \widehat{G}_i$ for $1 \leq i \leq k$. Note that any isomorphism from $\widehat{G} - E_i$ to $\widehat{G}_i$ must map $v_0$ to $v_0$, since the degree of $v_0$ in $\widehat{G} - E_i$ is at most $n$, and the degree of all vertices $v \neq v_0$ in $\widehat{G}_i$ is greater than $n$. Since the degree of $v_0$ in $\widehat{G}_i$ is $n$, the degree of $v_0$ in $\widehat{G} - E_i$ is also $n$, and thus no edge in $E_i$ is incident on $v_0$. The isomorphism from $\widehat{G} - E_i$ to $\widehat{G}_i$ must map $V(G)$ to $V(G_i)$, since these are exactly the sets of vertices adjacent to $v_0$. From this, it is easy to see that no edge in $E_i$ can be incident on a vertex not in $V(G)$, i.e., all edges in $E_i$ occur in $G$. It follows that $G - E_i \cong G_i$. This implies that $[G_1, \ldots, G_k] \subseteq \text{edge-deck}_c(G)$, and thus, $\langle G, [G_1, \ldots, G_k] \rangle \in k\text{-EDC}_c$.

Note that $\lambda(\widehat{G}) > c$ and $\|V(\widehat{G})\| \geq 4$. Thus, by Lemma 3.9, $\langle \widehat{G}, [\widehat{G}_1, \ldots, \widehat{G}_k] \rangle \in k\text{-EDC}_c \iff \langle L(\widehat{G}), [L(\widehat{G}_1), \ldots, L(\widehat{G}_k)] \rangle \in k\text{-VDC}_c$. It follows that $\sigma$ is a logspace many-one reduction from $k\text{-EDC}_c$ to $k\text{-VDC}_c$. ∎

**Lemma 3.11** *For all $c \geq 1$ and $k \geq 2$, $\text{GI} \leq_m^l k\text{-VDC}_c$ and $\text{GI} \leq_m^l k\text{-EDC}_c$.*

**Proof** Fix a $c \geq 1$ and a $k \geq 2$. By Lemma 3.10, it suffices to show that $\text{GI} \leq_m^l k\text{-EDC}_c$. Without loss of generality, we restrict to input instances $\langle G, H \rangle$ of GI such that $G$ and $H$ are connected graphs with at least three vertices. The reduction $\sigma$ from GI to $k\text{-EDC}_c$ is defined by $\sigma(\langle G, H \rangle) =$

$\langle G \cup cK_2,$
$[H \cup 2cK_1] \cup [\text{any } (k-1) \text{ cards from } (\text{edge-deck}_c(G \cup cK_2) - [G \cup 2cK_1])] \rangle.$

Using the techniques of the proof of [KH94, Lemma 3.2], it can be shown that $\sigma$ is a logspace many-one reduction from GI to $k\text{-EDC}_c$. Note also the similarity to the proof of Lemma 3.3. ∎

**Lemma 3.12** *For all $c \geq 1$ and $k \geq 2$, $k$-VDC$_c \leq_m^p$ GI and $k$-EDC$_c \leq_m^p$ GI.*

**Proof** By Corollary 2.3 and Lemma 3.10, it suffices to show that $k$-VDC$_c \leq_{dtt}^p$ GI. By definition, $[G_1, \ldots, G_k] \subseteq \text{vertex-deck}_c(G) \iff$

$$\bigvee_{[H_1,\ldots,H_k] \subseteq \text{vertex-deck}_c(G)} [H_1, \ldots, H_k] = [G_1, \ldots, G_k].$$

By the construction in Lemma 3.1 of [KH94], for each choice of $[H_1 \ldots, H_k] \subseteq$ vertex-deck$_c(G)$, there are graphs $\mathcal{H}_1$ and $\mathcal{H}_2$ such that $[G_1, \ldots, G_k] = [H_1, \ldots, H_k]$ if and only if $\langle \mathcal{H}_1, \mathcal{H}_2 \rangle \in$ GI. This completes the proof that $k$-VDC$_c \leq_{dtt}^p$ GI, since the "$\vee$" above is over $\binom{n}{\binom{n}{c}}_k$ instances of GI, where $n = ||V(G)||$, which is polynomial in $n$ for every fixed choice of $c$ and $k$.

In Theorem 3.13, we establish the polynomial-time isomorphism between GI and $k$-VDC$_c$, and between GI and $k$-EDC$_c$.

**Theorem 3.13** *For every $c \geq 1$ and $k \geq 2$, GI is polynomial-time isomorphic to $k$-VDC$_c$ and $k$-EDC$_c$.*

**Proof** Immediate from Lemma 3.11, Lemma 3.12 and the techniques used in [KH94, Lemma 5.6].

### 3.2.2 Legitimate Subdeck Problems

We now consider the relative complexity of GI and $k$-LVD$_c$, and that of GI and $k$-LED$_c$, for $k \geq 2$. Lemma 3.14 gives an alternate characterization of an instance of 2-LVD$_c$ in terms of polynomially many instances of GI. We will use Lemma 3.14 to obtain a polynomial-time many-one reduction from 2-LVD$_c$ to GI.

**Lemma 3.14** *For each $c \geq 1$, $[G_1, G_2]$ is a legitimate $c$-vertex-subdeck if and only if there exist $U_1 \subseteq V(G_1)$ and $U_2 \subseteq V(G_2)$, where $1 \leq ||U_1|| = ||U_2|| \leq c$, such that $G_1 - U_1$ is isomorphic to $G_2 - U_2$.*

**Proof** Fix a $c \geq 1$. Suppose that $[G_1, G_2]$ is a legitimate $c$-vertex-subdeck. By definition, there exist a graph $G$, distinct sets $T_1, T_2 \subseteq V(G)$, where $||T_1|| = ||T_2|| = c$, and isomorphisms $\psi_1$ from $G - T_1$ to $G_1$ and $\psi_2$ from $G - T_2$ to $G_2$. Clearly, $G_1 - \psi_1(T_2 - T_1)$ is isomorphic to $G - (T_1 \cup T_2)$ and $G_2 - \psi_2(T_1 - T_2)$ is isomorphic to $G - (T_1 \cup T_2)$. Thus, $G_1 - \psi_1(T_2 - T_1)$ is isomorphic to $G_2 - \psi_2(T_1 - T_2)$.

Now suppose that there exist $U_1 = \{u_{1,1}, u_{1,2}, \ldots, u_{1,\ell}\} \subseteq V(G_1)$ and $U_2 = \{u_{2,1}, u_{2,2}, \ldots, u_{2,\ell}\} \subseteq V(G_2)$, where $1 \leq \ell \leq c$, such that $G_1 - U_1$ is isomorphic to $G_2 - U_2$ via $\psi$. We now construct a graph $\mathcal{G}_2$ by adding new vertices $v_{2,1}, \ldots, v_{2,c}$ to $G_2$ and by including new edges incident on them. The graph $\mathcal{G}_2$ is defined as follows. Initially, $\mathcal{G}_2 := G_2$. For each $1 \leq i \leq \ell$, add a vertex $v_{2,i}$ to $\mathcal{G}_2$ and connect $v_{2,i}$ to every vertex in $\psi(N_{G_1}(u_{1,i}) - U_1)$. For each $1 \leq i < j \leq \ell$, add an edge $\{v_{2,i}, v_{2,j}\}$ to $\mathcal{G}_2$ if and only if $\{u_{1,i}, u_{1,j}\} \in E(G_1)$. Finally, for each $1 \leq i \leq c - \ell$, add a (isolated) vertex $v_{2,\ell+i}$ to $\mathcal{G}_2$. We construct another graph $\mathcal{G}_1$ in a similar way. Initially, $\mathcal{G}_1 := G_1$. For each $1 \leq i \leq \ell$, add a vertex $v_{1,i}$ to $\mathcal{G}_1$ and connect $v_{1,i}$ to every vertex in $\psi^{-1}(N_{G_2}(u_{2,i}) - U_2)$. For each $1 \leq i < j \leq \ell$, add an edge $\{v_{1,i}, v_{1,j}\}$ to $\mathcal{G}_1$ if and only if $\{u_{2,i}, u_{2,j}\} \in E(G_2)$. Finally, for each $1 \leq i \leq c - \ell$, add a (isolated) vertex $v_{1,\ell+i}$ to $\mathcal{G}_1$.

Let $\psi' : V(\mathcal{G}_1) \to V(\mathcal{G}_2)$ be defined as follows: $\psi'(v) = \psi(v)$ for all $v \in V(G_1 - U_1)$, for every $1 \leq i \leq \ell$, $\psi'(u_{1,i}) = v_{2,i}$ and $\psi'(v_{1,i}) = u_{2,i}$, and for every $\ell+1 \leq i \leq c$, $\psi'(v_{1,j}) = v_{2,j}$. It can be verified that $\psi'$ is an isomorphism from $\mathcal{G}_1$ to $\mathcal{G}_2$. Since $G_1$ ($= \mathcal{G}_1 - \{v_{1,1}, \ldots, v_{1,c}\}$) is a $c$-vertex-card of $\mathcal{G}_1$ and $G_2$ ($= \mathcal{G}_2 - \{v_{2,1}, \ldots, v_{2,c}\}$) is a $c$-vertex-card of $\mathcal{G}_2$, and since $\{v_{2,1}, \ldots, v_{2,c}\} \neq \psi'(\{v_{1,1}, \ldots, v_{1,c}\})$, it follows that $[G_1, G_2]$ is a legitimate $c$-vertex-subdeck. ∎

**Corollary 3.15** *For every $c \geq 1$, 2-LVD$_c \leq^p_m$ GI.*

**Proof** From Lemma 3.14, 2-LVD$_c \leq^p_{dtt}$ GI. By Corollary 2.3, it follows that 2-LVD$_c \leq^p_m$ GI. ∎

**Lemma 3.16** *For every $c \geq 1$ and $k \geq 2$, GI $\leq^l_m$ $k$-LVD$_c$.*

**Proof** Fix a $c \geq 1$ and a $k \geq 2$. Without loss of generality, we restrict to input instances $\langle G, H \rangle$ of GI where both $G$ and $H$ are connected graphs on $n > c$ vertices. Let $\ell = n+k$. We define a logspace many-one reduction $\sigma$ from this input-restricted version of GI to $k$-LVD$_c$ as follows:

$$\sigma(\langle G, H \rangle) = [\underbrace{K_\ell \cup K_{\ell+2c} \cup G, \ldots, K_\ell \cup K_{\ell+2c} \cup G}_{(k-1) \text{ copies}}, K_{\ell+c} \cup K_{\ell+c} \cup H].$$

Clearly, if $G$ and $H$ are isomorphic then $K_{\ell+c} \cup K_{\ell+2c} \cup G$ is a $c$-vertex-preimage of $\sigma(\langle G, H \rangle)$. Now suppose that $\sigma(\langle G, H \rangle) \in k$-LVD$_c$. Call $K_\ell \cup K_{\ell+2c} \cup G$ the $G$-card and call $K_{\ell+c} \cup K_{\ell+c} \cup H$ the $H$-card. A $c$-vertex-preimage of $\sigma(\langle G, H \rangle)$ can be obtained by adding $c$ vertices $v_1, \ldots, v_c$ and edges incident on them to the $G$-card. This $c$-vertex-preimage can also be obtained from the $H$-card by adding $c$ vertices $w_1, \ldots, w_c$ and edges incident on them to the $H$-card. It is immediate that every $v_i$ is connected to every element of $K_\ell$ and that all the $v_i$'s are connected to each other in the $c$-vertex-preimage obtained from the $G$-card. Similarly, all the $w_i$'s are connected to each other and all the $w_i$'s are connected to all the vertices in the same $K_{\ell+c}$ in the $c$-vertex-preimage obtained from the $H$-card.

It follows that in the $c$-vertex-preimage obtained from the $G$-card, the vertices in $V(G)$ are exactly the vertices of degree $\leq n + c$. Likewise, in the $c$-vertex-preimage obtained from the $H$-card, the vertices in $V(H)$ are exactly the vertices of degree $\leq n + c$. It follows that the preimages obtained from the $G$-card and the $H$-card are isomorphic only if $G \cong H$. ∎

From Corollary 3.15 and Lemma 3.16, we get that, for each $c \geq 1$, GI $\equiv_m^p$ 2-LVD$_c$. We further note that GI is polynomial-time isomorphic to 2-LVD$_c$.

**Theorem 3.17** *For every $c \geq 1$, GI is polynomial-time isomorphic to 2-LVD$_c$.*

**Proof** Proof omitted. Refer to the proof of Theorem 3.20 for the technique. ∎

**Lemma 3.18** *For every $c \geq 1$ and $k \geq 2$, $k$-LED$_c$ $\leq_m^p$ GI.*

**Proof** The exact same construction as in Lemma 3.4 shows that $k$-LED$_c$ $\leq_{dtt}^p$ $k$-EDC$_c$. Thus, by Lemma 3.12 and Corollary 2.3, $k$-LED$_c$ $\leq_m^p$ GI. ∎

**Lemma 3.19** *For every $c \geq 1$ and $k \geq 2$, GI $\leq_m^l$ $k$-LED$_c$.*

**Proof** Fix a $c \geq 1$ and a $k \geq 2$. Without loss of generality, let $G$ and $H$ be connected graphs on $n$ vertices and let $n > c$. Let $\ell = n + k$. For $i = \ell, \ell + 1$, let $S_{i,j}$, where $1 \leq j \leq \binom{\binom{i}{2}}{c}$, be an enumeration of sets of $c$ distinct edges of $K_i$. The logspace many-one reduction function $\sigma$ is defined by $\sigma(\langle G, H \rangle) =$

$$[G \cup (K_\ell - S_{\ell,1}) \cup K_{\ell+1}, \ldots, G \cup (K_\ell - S_{\ell,k-1}) \cup K_{\ell+1}, H \cup K_\ell \cup (K_{\ell+1} - S_{\ell+1,1})].$$

First note that if $G \cong H$, then $G \cup K_\ell \cup K_{\ell+1}$ is a valid $c$-edge-preimage of $\sigma(\langle G, H \rangle)$. For the converse, suppose that $\sigma(\langle G, H \rangle) \in k$-LED$_c$. Call $G \cup (K_\ell - S_{\ell,1}) \cup K_{\ell+1}$ the $G$-card and $H \cup K_\ell \cup (K_{\ell+1} - S_{\ell+1,1})$ the $H$-card.

Note that the only way for a $c$-edge-preimage obtained from the $H$-card to include $K_{\ell+1}$ as a subgraph is to add $c$ edges to $(K_{\ell+1} - S_{\ell+1,1})$ because it takes $\ell$ edges to completely connect a vertex to $K_\ell$ and $\ell > c$. It follows that the only possible preimage (up to isomorphism) is the graph $H \cup K_\ell \cup K_{\ell+1}$. If this graph is a $c$-edge-preimage of the $G$-card, the only way to turn the $G$-card into the $c$-edge-preimage is to add the $c$ missing edges to $K_\ell - S_{\ell,1}$. It follows that $G \cong H$. ∎

Lemma 3.18 and Lemma 3.19 imply that GI $\equiv_m^p$ $k$-LED$_c$. In Theorem 3.20, we strengthen the polynomial-time many-one equivalence of GI and $k$-LED$_c$ to their polynomial-time isomorphism.

**Theorem 3.20** *For every $c \geq 1$ and $k \geq 2$, GI is polynomial-time isomorphic to $k$-LED$_c$.*

**Proof** Fix a $c \geq 1$ and a $k \geq 2$. Since GI $\equiv_m^p$ $k$-LED$_c$ (from Lemma 3.18 and 3.19), it suffices to show (by Theorem 2.8 and Lemmas 2.9 and 2.10) that $k$-LED$_c$ has polynomial-time computable functions $S_{k\text{-LED}_c}$ and $D_{k\text{-LED}_c}$ satisfying Lemma 2.5. Let $[G_1, \ldots, G_k]$ be an instance of $k$-LED$_c$ and let $y \in \Sigma^*$. The function $S_{k\text{-LED}_c}$ is defined as follows. On input $([G_1, \ldots, G_k], y)$,

1. Compute the polynomial-time many-one reduction from $k$-LED$_c$ to GI of Lemma 3.18 on input $[G_1, \ldots, G_k]$ and let $\langle \mathcal{H}_1, \mathcal{H}_2 \rangle$ be the output of the reduction.

2. Compute $S_{\text{GI}}(\langle \mathcal{H}_1, \mathcal{H}_2 \rangle, y)$ of Lemma 2.9 and let $\langle \widehat{\mathcal{H}}_1, \widehat{\mathcal{H}}_2 \rangle$ be the output of this step.

3. Compute the logspace many-one reduction from GI to $k$-LED$_c$ of Lemma 3.19 on input $\langle \widehat{\mathcal{H}}_1 + K_{c+1}, \widehat{\mathcal{H}}_2 + K_{c+1} \rangle$. Output the string computed in this step.

From the definition of many-one reducibility and that of $S_{\text{GI}}$, it follows that $S_{k\text{-LED}_c}$ is a polynomial-time computable function satisfying Lemma 2.5. We now define $D_{k\text{-LED}_c}$ in terms of the output of a polynomial-time transducer that works as follows. On input $[H_1, \ldots, H_k]$,

1. Scan $[H_1, \ldots, H_k]$ to find an edge-card of the form $(\widehat{\mathcal{H}}_1 + K_{c+1}) \cup (K_\ell - S) \cup K_{\ell+1}$, where $\ell = ||V(\widehat{\mathcal{H}}_1)|| + c + 1 + k$, $S \subseteq E(K_\ell)$, and $||S|| = c$. If no such card exists, then output the string "undefined."

2. Scan $[H_1, \ldots, H_k]$ to find an edge-card of the form $(\widehat{\mathcal{H}}_2 + K_{c+1}) \cup K_\ell \cup (K_{\ell+1} - S')$, where $\ell = ||V(\widehat{\mathcal{H}}_2)|| + c + 1 + k$, $S' \subseteq E(K_{\ell+1})$, and $||S'|| = c$. If no such card exists, then output the string "undefined."

3. Output $D_{\text{GI}}(\langle \widehat{\mathcal{H}}_1, \widehat{\mathcal{H}}_2 \rangle)$, where $D_{\text{GI}}$ is the polynomial-time computable function of Lemma 2.9.

By the construction in Lemma 3.19 and the definition of $D_{\text{GI}}$, it follows easily that $D_{k\text{-LED}_c}$ satisfies Lemma 2.5. ∎

# 4 Reconstruction Numbers of Graphs

The ally-reconstruction number [HP85,Myr89] of a graph $G$ is the minimum number of one-vertex-deleted subgraphs (i.e., 1-vertex-cards) that identify $G$ (up to isomorphism). Since the ally-reconstruction number of a graph $G$ is characterized by the *existence* of some set of that number of 1-vertex-cards of $G$ that identify $G$, we will denote this number by $vrn_\exists(G)$. Likewise, we use $ern_\exists(G)$ to denote the minimum number of 1-edge-cards that identify $G$. We also define an analogous concept of reconstruction number for a graph $G$, denoted by $vrn_\forall(G)$ ($ern_\forall(G)$), in which a certain number of 1-vertex-cards (1-edge-cards) of $G$, irrespective of their choice, will always suffice to recognize $G$. Thus, no matter which 1-vertex-cards (1-edge-cards) an adversary selects from the deck of $G$, $vrn_\forall(G)$ ($ern_\forall(G)$) many 1-vertex-cards (1-edge-cards) are enough to identify $G$ up to isomorphism. If such a number doesn't exist, we define it to be $\infty$. Formal definitions of the various reconstruction numbers are as follows.

**Definition 4.1** *For any graph $G$,*

1. *$vrn_\exists(G)$ (known as the* ally-reconstruction number *[HP85,Myr89]) is the minimum number such that there is a collection of $vrn_\exists(G)$ 1-vertex-cards of $G$ that identify $G$ (up to isomorphism). If this number does not exist, then $vrn_\exists(G) = \infty$.*

2. *$ern_\exists(G)$ is the minimum number such that there is a collection of $ern_\exists(G)$ 1-edge-cards of $G$ that identify $G$ (up to isomorphism). If this number does not exist, then $ern_\exists(G) = \infty$.*

3. *$vrn_\forall(G)$ is the minimum number such that every collection of $vrn_\forall(G)$ 1-vertex-cards of $G$ identify $G$ (up to isomorphism). If this number does not exist, then $vrn_\forall(G) = \infty$.*

4. *$ern_\forall(G)$ is the minimum number such that every collection of $ern_\forall(G)$ 1-edge-cards of $G$ identify $G$ (up to isomorphism). If this number does not exist, then $ern_\forall(G) = \infty$.*

It is clear that for any graph $G$ for which $vrn_\exists(G) < \infty$ ($ern_\exists(G) < \infty$), $vrn_\exists(G) \leq vrn_\forall(G) \leq ||V(G)||$ ($ern_\exists(G) \leq ern_\forall(G) \leq ||E(G)||$). Note that $vrn_\exists(G)$ is finite for every graph $G$ having at least three vertices if and only if the Reconstruction Conjecture is true, and $ern_\exists(G)$ is finite for every graph $G$ having at least four edges if and only if the Edge-Reconstruction Conjecture is true. Theorem 4.2 says that for any disconnected graph $G$ having at least three vertices, $vrn_\exists(G)$ is finite (from which one can conclude that $vrn_\forall(G)$ is finite, since certainly choosing to include all cards will include whichever particular $vrn_\exists(G)$ cards were already enough to determine $G$, and so certainly $vrn_\forall(G) \leq ||V(G)||$).

**Theorem 4.2 ([Myr89]; proof corrections in [Mol95])** *If $G$ is a disconnected graph with not all components isomorphic then $vrn_\exists(G) = 3$. Moreover, if $G$ has at least three vertices and is a disconnected graph with all components isomorphic then $vrn_\exists(G) \leq c + 2$ where $c$ is the number of vertices in a component.*

In the next lemma, we give an example of a family of disconnected graphs $G$ (parameterized by $n$, the number of vertices of the $n$'th graph in the family) for which $vrn_\exists(G) < vrn_\forall(G)$.

**Lemma 4.3** *For all $n \geq 4$, there exists a disconnected graph $G_n$ such that $||V(G_n)|| = n$ and $vrn_\exists(G_n) < vrn_\forall(G_n)$.*

**Proof** Let $n \geq 4$. Let $n = 2t$ if $n$ is even, and let $n = 2t+1$ if $n$ is odd. Define the ordered pair

$$(G_n, H_n) = \begin{cases} (K_{t+1} \cup K_{t-1}, 2K_t) & \text{if } n \text{ is even,} \\ (K_{t+1} \cup K_{t-1} \cup K_1, 2K_t \cup K_1) & \text{if } n \text{ is odd.} \end{cases}$$

By Theorem 4.2, $vrn_\exists(G_n) = 3$. It is clear that $G_n$ and $H_n$ are nonisomorphic graphs. For even $n$, both $G_n$ and $H_n$ have $t+1$ 1-vertex-cards that are isomorphic to $K_t \cup K_{t-1}$, and for odd $n$, both $G_n$ and $H_n$ have $t+1$ 1-vertex-cards that are isomorphic to $K_t \cup K_{t-1} \cup K_1$. Thus, $vrn_\forall(G_n) \geq t + 2 > 3 = vrn_\exists(G_n)$.

Actually, one can show that we have the equality $vrn_\forall(G_n) = vrn_\forall(H_n) = t + 2$, for all $n \geq 4$. It remains to show that $t+2$ is an upper bound in all four cases, i.e., $n$ odd or even and $G_n$ or $H_n$.

$vrn_\forall(H_{2t}) \leq t + 2$ follows from Theorem 4.3, since all cards in vertex-deck$_1(H_{2t})$ are isomorphic.

In the analysis of the three remaining cases we will use the following fact.

**Fact 4.4** *If vertex-deck$_1(G)$ has 4 cards all of whose components are complete graphs, then all components of $G$ are also complete graphs.*

**Proof of Fact 4.4** Let $x_1$, $x_2$, $x_3$, and $x_4$ be four distinct vertices in $G$ such that for all $i$, $1 \leq i \leq 4$, all components of $G - x_i$ are complete graphs. We employ proof by contradiction.

In particular, let us suppose that $u$ and $v$ are two vertices in $G$ such that there exists a path from $u$ to $v$ in $G$ and $\{u,v\} \notin E(G)$. Note that $x_1$, $x_2$, $x_3$, and $x_4$ must lie on each path from $u$ to $v$ (possibly as endpoints), since $u$ and $v$ are not in the same connected component of $G - x_i$ for $1 \leq i \leq 4$. Without loss of generality, there exists a simple path from $u$ to $v$ that goes from $u$ to $x_1$ (it is possible that $u = x_1$), from $x_1$ to $x_2$, from $x_2$ to $x_3$, from $x_3$ to $x_4$, and from $x_4$ to $v$ (it is possible that $v = x_4$). Now consider the card $G - x_4$. Clearly, $x_1$, $x_2$, and $x_3$ occur in the same connected component of $G - x_4$. It follows that $\{x_1, x_3\} \in E(G)$. But then $u$ and $v$ occur in the same connected component of $G - x_2$, which implies that $\{u,v\} \in E(G)$. This is a contradiction.  ■ (Fact 4.4)

All components of all cards in the vertex-decks of $G_n$ and $H_n$ are complete graphs, and $n \geq 4$, so in the following we will assume that all preimage graphs considered are disjoint unions of complete graphs. By Fact 4.4 we only need to show how to reconstruct the orders of components from any subdeck of $t+2$ cards, $t \geq 2$.

$G = G_{2t}$. The deck of $G$ contains cards of two types, with possible orders of components $T_1 = (t+1, t-2)$ (if $t = 2$, we identify $(t+1, t-2)$ with $(t+1)$, etc.) and $T_2 = (t, t-1)$ and has $t-1$ and $t+1$ of such cards, respectively. The possible preimages of cards of type $T_1$ have component orders $(t+1, t-2, 1)$, $(t+2, t-2)$ or $(t+1, t-1)$, and those of type $T_2$ have orders $(t, t)$, $(t, t-1, 1)$ or $(t+1, t-1)$. Note that the only possible preimage of two cards of different types has orders $(t+1, t-1)$, which is that of $G$. $vrn_\forall(G_{2t}) \leq t+2$ follows, since any $t+2$ cards must contain a card of type $T_1$ and a card of type $T_2$.

$G = G_{2t+1}$. If $t > 2$, there are three types of component orders of cards in the deck: $T_1 = (t, t-1, 1)$, $T_2 = (t+1, t-2, 1)$ or $T_3 = (t+1, t-1)$, and there are $t+1$, $t-1$ and $1$ of such cards, respectively. If $t = 2$, there are only two types of component orders, since $T_2$ and $T_3$ are the same type. Given $t+2$ cards, one of the cards will be of type $T_1$ and one will be of type $T_2$ or $T_3$. The only preimage of a card of type $T_3$ which is also possible for $T_1$ has the order type $(t+1, t-1, 1)$ of $G$ itself, and thus a card of type $T_1$ together with a card of type $T_3$ reconstruct orders of components of $G$. If $t > 2$ and we have one card of type $T_1$ and one card of type $T_2$, then a case analysis similar to that of $G = G_{2t}$ completes the proof that $vrn_\forall(G_{2t+1}) \leq t+2$.

$G = H_{2t+1}$. The proof in this case follows again the same template. There are two types of orders of components of cards in the deck: $T_1 = (t, t-1, 1)$, $T_2 = (t, t)$, and there are $2t$ and $1$ of such cards, respectively. The only preimage of a card of type $T_2$ which is possible for $T_1$ has the order type $(t, t, 1)$ of $G$ itself, and thus a card of type $T_2$ together with a card of type $T_1$ reconstruct orders of components of $G$. If $t > 2$, the possible preimages of orders of type $T_1$ are $(t, t-1, 1, 1)$, $(t+1, t-1, 1)$, $(t, t-1, 2)$ and $(t, t, 1)$ and in turn such graphs have in their decks 2, $t+1$, 1, and $2t$ cards of type $T_1$, respectively. Thus if all $t+2$ cards have order type $T_1$, then all must come from the preimage of type $(t, t, 1)$. If $t = 2$, the possible preimages of orders of type $T_1$ are $(2, 1, 1, 1)$, $(3, 1, 1)$, and $(2, 2, 1)$ and in turn such graphs have in their decks 3, 3, and 4 cards of type $T_1$, respectively. Hence $vrn_\forall(G_{2t+1}) \leq t+2$. ∎

The Reconstruction Conjecture can be restated as follows: For each $n \geq 3$, given any collection $\mathcal{D}$ of $n$ graphs with $n-1$ vertices in each, there can be at most one 1-vertex-preimage of $\mathcal{D}$. What can we say about the number of nonisomorphic 1-vertex-preimages of a collection $\mathcal{D}$ of graphs with $n-1$ vertices in each where the size of $\mathcal{D}$ is smaller than $n$? Myrvold [Myr90] showed that for any tree $T$, the number of nonisomorphic preimages of endvertex-deck($T$) is exactly one; the unique preimage up to isomorphism is $T$ itself. However, the following theorem by Bryant [Bry71] says that there are graphs $G$ for which the endvertex-deck($G$) has more than one nonisomorphic preimage.

**Theorem 4.5 ([Bry71])** *For any positive integer $k$, there exist nonisomorphic graphs $G$ and $H$, with $k$ endvertices in each, such that* endvertex-deck($G$) = endvertex-deck($H$).

Note that Theorem 4.5 claims only the existence of at least two nonisomorphic 1-vertex-preimages of a certain collection consisting of $k$ 1-vertex-cards, for every $k \geq 2$. In the next theorem, we show that there is a family of multisets of $k$ graphs on $(2^{k-1}+1)n+k$ vertices with $2^n$ 1-vertex-preimages.

**Theorem 4.6** *For all $k \geq 2$ and $n \geq 1$, there is a deck $\mathcal{D}$ of $k$ 1-vertex-cards on $(2^{k-1}+1)n+k$ vertices with at least $2^n$ 1-vertex-preimages.*

**Proof** Each of the $k$ 1-vertex-cards in $\mathcal{D}$ is identical, and defined as follows.

1. $x_0, \ldots, x_n$ are the vertices of the path graph $P_{n+1}$ ($\{x_i, x_{i+1}\}$ is an edge for $0 \leq i \leq n$).

2. $y_1, \ldots, y_{k-1}$ are special selector vertices.

3. For $i := 1 \ldots n$,

   **3.1** Let $G_i$ be the complete graph $K_{2^{k-1}}$ and let $V(G_i) = \{z_{i,Y} \mid Y \subseteq \{y_1, \ldots, y_{k-1}\}\}$.

   **3.2** Connect $x_i$ to all the vertices of $G_i$.

   **3.3** For each $z_{i,Y} \in V(G_i)$, connect $z_{i,Y}$ to each vertex $y$ such that $y \in Y$.

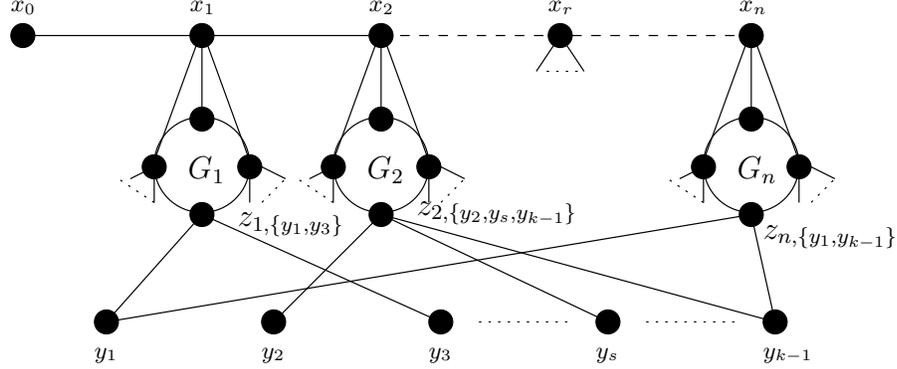

Figure 1: A 1-vertex-card in $\mathcal{D}$

Consider 1-vertex-preimages $H$ of $\mathcal{D}$ of the following form.

1. $x_0, \ldots, x_n$ are the vertices of the path graph $P_{n+1}$ ($\{x_i, x_{i+1}\}$ is an edge for $0 \leq i \leq n$).

2. $y_1, \ldots, y_{k-1}, y_k$ are special selector vertices.

3. For $i := 1 \ldots n$,

3.1 Let $G_i$ be the complete graph $K_{2^{k-1}}$ and let $V(G_i) = \{z_{i,j} \mid j \in \{1, \ldots, 2^{k-1}\}\}$.

3.2 Connect $x_i$ to all the vertices of $G_i$.

3.3 The edges between the $y$-vertices and $G_i$ are defined according to one of the following two cases.

   **Case 1.** Let $Y_{i,1}, \ldots, Y_{i,2^{k-1}}$ be an enumeration of the subsets of $\{y_1, \ldots, y_k\}$ of odd size. For each $j \in \{1, \ldots, 2^{k-1}\}$, connect $z_{i,j}$ to each vertex $y$ such that $y \in Y_{i,j}$.

   **Case 2.** Let $Y_{i,1}, \ldots, Y_{i,2^{k-1}}$ be an enumeration of the subsets of $\{y_1, \ldots, y_k\}$ of even size. For each $j \in \{1, \ldots, 2^{k-1}\}$, connect $z_{i,j}$ to each vertex $y$ such that $y \in Y_{i,j}$.

Figure 1 shows the construction of a 1-vertex-card in $\mathcal{D}$. Note that $H$ is a 1-vertex-preimage of $\mathcal{D}$, since $H - y_\ell$ is isomorphic to the 1-vertex-card in $\mathcal{D}$ for $1 \leq \ell \leq k$ (via an isomorphism $\pi$ that maps $x_i$ to $x_i$ for $1 \leq i \leq n$, $\{y_1, \ldots, y_k\} - \{y_\ell\}$ to $\{y_1, \ldots, y_{k-1}\}$ (arbitrarily), and $z_{i,j}$ to $z_{i,\pi[Y_{i,j} - \{y_\ell\}]}$ for $1 \leq i \leq n, 1 \leq j \leq 2^{k-1}$).

As $i$ varies from 1 to $n$, in step 3.3 each time we can apply either Case 1 or Case 2. Every two distinct sequences of such choices in the construction of $H$ give rise to nonisomorphic graphs. Thus, the number of nonisomorphic 1-vertex-preimages is at least $2^n$. ∎

## 5 Open Problems

In this section, we mention some open problems. Theorem 3.17 states that, for every $c \geq 1$, GI $\equiv_{iso}^{p}$ 2-LVD$_c$. However, for $k > 2$ and $c \geq 1$, we do not know whether $k$-LVD$_c$ is polynomial-time equivalent to GI or is NP-complete (or is neither). Since for $k > 2$ and $c \geq 1$ it is not clear, even under the assumption that the Reconstruction Conjecture is true, whether $k$-LVD$_c$ is low for PP, it is at least possible that $k$-LVD$_c$ is NP-complete.

It also would be interesting to investigate the complexity of problems related to the reconstruction numbers. So we define the following problems:

1. EXIST-VRN= $\{\langle G, k \rangle \mid vrn_\exists(G) \leq k\}$.

2. UNIV-VRN= $\{\langle G, k \rangle \mid vrn_\forall(G) \leq k\}$.

3. EXIST-ERN= $\{\langle G, k \rangle \mid ern_\exists(G) \leq k\}$.

4. UNIV-ERN= $\{\langle G, k \rangle \mid ern_\forall(G) \leq k\}$.

It is easy to see that EXIST-VRN $\in \Sigma_2^p$ (since GI is low for $\Sigma_2^p$), UNIV-VRN $\in$ coNP$^{\text{GI}}$, EXIST-ERN $\in$ NP$^{\text{GI}}$ (we get a better upper bound than for EXIST-VRN since, as in the proof of Lemma 3.4, we have to consider only polynomially many possible preimages of the edge-deck), and UNIV-ERN $\in$ coNP$^{\text{GI}}$. It would be interesting to obtain tight (or tighter) bounds on the complexity of these problems. For instance, is EXIST-VRN complete for $\Sigma_2^p$? Is UNIV-ERN coNP-hard?

**Acknowledgments** We thank the anonymous referees for helpful comments.